\documentclass[preprint,epsf,aps,floats]{revtex4}
\usepackage{amsmath,amssymb,bm,epsfig,psfrag}
\usepackage{mathtools}

\DeclarePairedDelimiter\abs{\lvert}{\rvert}%

\newcommand\+{\dagger}

\newcommand{\be}{\begin{equation}}
\newcommand{\ee}{\end{equation}}
\newcommand{\ber}{\begin{eqnarray}}
\newcommand{\eer}{\end{eqnarray}}

\newcommand\bra[1]{\langle #1 |}
\newcommand\ket[1]{|{#1}\rangle}

\def\Dsl{\,\raise.15ex \hbox{/}\mkern-12.8mu D}

\begin{document}
%\preprint{}

\title{Phonon-mediated relaxation in nanomaterials from combining Density Functional Theory based 
non-adiabatic molecular dynamics with Kadanoff-Baym-Keldysh technique} 

\author{Hadassah~B.~Griffin}
\affiliation{Department of Physics,~North Dakota State University,~Fargo, ND~58108,~USA}

\author{Andrei~Kryjevski}
\affiliation{Department of Physics,~North Dakota State University,~Fargo, ND~58108,~USA}

\begin{abstract}
Boltzmann transport equation (BE) is a potent approach to dynamics of 
a photoexcited (nano)material. BE collision integrals for different 
relaxation channels can be systematically computed using the Kadanoff-Baym-Keldysh (KBK) 
formalism (also called NEGF) utilizing the Density Functional Theory (DFT) simulation output. 
However, accurate description of phonon-mediated relaxation in a general class of 
(nano)materials that includes exciton effects is still an outstanding problem. 
The approach proposed here is based on the observation that the non-adiabatic couplings 
of the DFT-based non-adiabatic molecular dynamics (NAMD) play the role of a 
time-dependent external potential coupled to the electrons. This allows application of the 
Keldysh approach resulting in the exciton-phonon BE collision integral, which incorporates 
exciton wave functions and energies obtained from Bethe-Salpeter equation. As an 
application, we augment BE with radiative recombination and photon-mediated exciton-exciton 
transition terms and then use it to calculate photoluminescence (PL) spectrum for several 
1.5-$nm$ semiconductor chalcogenide nanocrystals, such as $Cd_{37}Pb_{31}Se_{68},~Cd_{31}Pb_{37}Se_{68},$ 
which are Janus-type, and for $Pb_{68}Se_{68}.$  
\end{abstract}
\date{\today}
%\pacs{03.75.Ss}

\maketitle

%\section{Introduction}

%\section{Theoretical Methods and Approximations}
%\label{sec:method}

%\section{Conclusions and Outlook}

%\begin{thebibliography}{99}
%\end{thebibliography}
%\bibliographystyle{jpc}
%\bibliography{dftNSF2012}% Produces the bibliography via BibTeX.

%%%%%%%%%%%%%%%%%%%%%%%%%%%%%%%%%%%%%%%%%%%%%%%%%%%%%%%%%%%%%%%%%%%%%%%%%%%%%%%%%
\noindent
\section{Introduction}
The ability to describe relaxation processes in a photoexcited (nano)material 
%with which photoexcited charge carriers in a nanoparticle lose their energy to heat 
is important for its potential technological applications. For example, fast carrier 
relaxation is desirable in lasers and other light emitting devices, but in, {\it e.g.}, 
photovoltaics one aims to minimize the carrier heat loss rate as it limits the solar cell 
efficiency \cite{10.1063/1.1736034}. It is understood that this description has to be 
comprehensive so that different time-evolution channels, such as phonon-mediated 
carrier relaxation, energy transfer, recombination, both radiative and non-radiative, 
carrier multiplication, {\it etc.} are allowed to ``compete" with each other. Boltzmann 
transport equation (BE) is a potent approach to this problem \cite{Landau10}. 

On the other hand, computational studies of atomistic models of (nano)materials using 
{\it ab initio} electronic structure techniques, such as Density Functional Theory (DFT) 
\cite{PhysRev.136.B864,PhysRev.140.A1133}, have proven to be a helpful alternative to the
actual experiments. The results of these studies often serve as a unique source of insight 
into the system's properties. Note that using atomistic models naturally allows inclusion 
of surfaces, interfaces, dopants, ligands, {\it etc.}
since in DFT the defects and the bulk atoms are treated equally.
%\cite{Kilina_ACSnano12,Kilina_Nanoscale-12,Kilina_JACS09_057408}.
This justifies additional computational expense of DFT as compared to
other approaches to the electronic sructure, such as the effective mass
approximation, tight-binding approximation, {\it etc.} \cite{ashcroft1976solid,Kang:97}.

For the excited state description currently
%the approach that 
the most efficient and comprehensive {\it ab~initio} approach is the DFT-based 
many-body perturbation theory (MBPT-DFT), which stems from the formally exact Hedin 
equations for the correlation functions accompanied by the necessary truncations
and approximations \cite{PhysRev.139.A796}. 
In practice MBPT is a set of many-body 
%perturbative quantum mechanics 
techniques for various quantities                                                
%balances accuracy and applicability 
utilizing DFT output. Among them is the $GW$ method for 
single-particle energies.
%approximated from DFT 
%to the first order in the screened Coulomb interaction, while 
Bethe-Salpeter equation (BSE) is solved for the exciton states; 
a technique for the three-body states - trions - has also been developed~
\cite{PhysRevB.34.5390,PhysRevB.62.4927,PhysRevLett.116.196804,0034-4885-61-3-002,RevModPhys.74.601}. 
%%%%% 
In all these methods random phase approximation (RPA) is used for screening (see, {\it e.g.},\cite{FW}). 
%Currently, computational expense of MBPT scales as ${\cal O}(N^4),$ where 
%$N$ is the number of electrons \cite{PhysRevB.87.155148,Deslippe_2012}, although the 
%so-called stochastic $GW$ and BSE have better scaling \cite{PhysRevLett.113.076402,PhysRevB.91.235302}. 
The resulting energy levels and wave functions
%using RPA-screened Coulomb interactions 
are subsequently used in calculations of various excited-state properties 
({\it e.g}, \cite{PhysRevLett.80.3320,PhysRevLett.92.077402,doi:10.1021/acs.jpclett.8b02288}).

The finite-temperature weak-nonequlibrium extension of MBPT-DFT - the Kadanoff-Baym-Keldysh (KBK) 
technique \cite{Landau10} - has been applied extensively to the description of (nano)materils. 
In particular, it has been used to compute collision integrals in BE \cite{doi:10.1063/1.4997048}. 
In this method, the collision integrals are obtained from the self-energies computed to a 
given order in perturbation theory \cite{Landau10}. This approach has been used to include phonon-mediated 
relaxation including one- and two-phonon exchange processes, exciton transfer and carrier multiplication 
into BE studies of different photoexcited nanomaterials  \cite{doi:10.1063/1.4997048,doi:10.1021/acs.jpclett.8b02288,SFJCP}. 

However, the MBPT-DFT approaches to the phonon-mediated relaxation are based on 
perturbation theory in the coupling of electrons to the individual phonon modes
with the phonons in the harmonic approximation. Therefore, they
%approach is similar to the frozen phonon approximation and is 
are only applicable to rigid systems at low temperatures where atoms undergo small 
oscillations about their equilibrium positions \cite{vanCamp,PhysRevB.85.235422}.
%; the electronic states are approximated using 
%equilibrium atomic positions ${\bf R}_I^0$ \cite{vanCamp,PhysRevB.85.235422}.  

On the other hand, the time-domain DFT (tdDFT) is a well established nonperturbative approach that 
exploits coupling of electronic and ionic subsystems in the DFT-based non-adiabatic molecular dynamics 
(NAMD) at finite temperature. Here one uses a DFT program ({\it e.g.}, VASP, Quantum Espresso, {\it etc.}), 
to simulate thermal motion of the ions. Following along the nuclear MD trajectory 
one estimates the electron state response to the changes in the ion positions. This allows one 
to determine the nonadiabatic couplings (NACs) \cite{ISI:A1990DN23400021,ISI:A1994PF94300022}. 
NACs obtained from NAMD are used in different variants of this approach, such as the surface 
hopping techniques  \cite{PhysRevLett.95.163001}, on-the-fly techniques 
\cite{Han16022015,doi:10.1021/acs.jpcc.5b06434}, Redfield tensor approaches \cite{doi:10.1021/ct5004093}, 
machine learning \cite{doi:10.1021/acs.jpclett.8b02469,doi:10.1021/acs.jpclett.2c03097}. 
Work to include excitonic effects into the surafce-hopping approach has started \cite{jiang2021real}.
%%% from NSF16 grant
%An alternative approach for computing non-adiabatic couplings is the “on-the-fly” procedure.
%Here one uses a DFT program, such as VASP, to simulate thermal motion of the
%ions. Following along the nuclear MD trajectory one estimates the KS state response to
%the changes in RI and determines the couplings ˜gij(t) = hi(RI(t))| d/dt |j(RI(t))i [139, 140].
%This approach is non-perturbative and does not rely on the harmonic approximation.
%[139] J. Tully. Molecular-dynamics with electronic-transitions. The Journal of Chemical Physics,93(2):1061–1071, 1990.
%[140] S. Hammes-Schiffer and J. Tully. Proton-transfer in solution - molecular-dynamics with quantum transitions. The Journal of Chemical Physics, 101(6):4657–4667, 1994.

It would be beneficial to incorporate NACs into the MBPT-DFT technique and be able to compute the 
exciton-phonon BE collision integral without having to rely on the perturbation theory in the 
electron-phonon mode coupling, thus extending applicability of the BE approach. More generally, 
it would make it possible to obtain contributions to the collision integrals from all the time evolution 
channels within the same MBPT paradigm. 
%thus resulting in a comprehensive description of time evolution. 
Therefore, the main goal of this work is to develop an MBPT-DFT method for the phonon-mediated 
relaxation of a photoexcited nanomaterial with excitons as the degrees of freedom, which utilizes 
NACs, and, therefore, does not rely on the perturbation theory in the coupling of electrons to the 
individual phonon modes.
%A. Kryjevski, D. Mihaylov, and D. Kilin, The Journal of Physical Chem. Letters, 9(19) (2018)
%A. Kryjevski, D.Mihaylov, S. Kilina, and D. Kilin, The J. of Chem. Physics, 147, 154106 (2017)

In Section 2 we review the basic ideas of tdDFT, recognize NACs as a time-dependent potential for electrons, 
describe inclusion of NACs into MBPT-DFT and present the expressions for the collision integrals for exciton-phonon 
processes, and, also, for the radiative recombination. In section 3 we outline the procedure for 
computing photoluminescence (PL) spectrum using BE solution. In section 4 present results of application to PL to 
several 1.5-$nm$ chalcogenide nanocrystals (NCs). The conclusions and outlook are presented in section 5. 

\section{Incorporating Non-Adiabatic Couplings into Keldysh Many-Body Perturbation Theory}
For completeness let us start from the hamiltonian for both 
nuclei and electrons, which are coupled to photons, which is 
%Dynamics of valence electrons and ions 
%at positions $\{ {\bf R}_I \}, ~I=1,..,{\rm N}_{ion},$ 
%is governed by the following Hamiltonian
\ber
&&{\rm H}=\int{\rm d}{\bf x}~\psi^{\+}_{\alpha}({\bf x})\left(-\frac{\hbar^{2}}{2m}\nabla^{2}+V_{eN}({\bf x},{\bf R})\right)\psi_{\alpha}({\bf x})+V_C+{\rm H}_{N}
+{\rm H}_{Ae}+{\rm H}_{A},
\label{Hmicro}
\eer
where
\ber
V_{eN}&=&\sum_{I=1}^{{\rm N}_{i}}~{\rm v}({\bf x},{\bf R}_I),
%~{\rm v}({\bf x},{\bf R}_I)=-\frac{{\rm Z}_Ie^2}{|{\bf x}-{\bf R}_I|},
%\simeq -\sum_{I=1}^{{\rm N}_{ion}}~\frac{{\rm Z}_I e^2}{|{\bf x}-{\bf R}_I|}, \\
%\frac{{\rm Z}_I e^2}{|{\bf x}-{\bf R}_I|}
\cr
V_C&=&\int{\rm d}{\bf x}{\rm d}{\bf y}~\psi^{\+}_{\alpha}({\bf x})\psi^{\+}_{\beta}({\bf y})\frac{e^2}{2|{\bf x}-{\bf y}|}\psi_{\beta}({\bf y})\psi_{\alpha}({\bf x}), 
%\nonumber 
\cr
{\rm H}_{N}&=&\sum_{I=1}^{{\rm N}_{i}}\frac{P_I^2}{2 M_I}+V_{NN},~V_{NN} \simeq \frac12\sum_{I\neq J}\frac{{\rm Z}_I{\rm Z}_J e^2}{|{\bf R}_I-{\bf R}_J|},
\cr
{\rm H}_{Ae}&=&-\frac{1}{c}\int{\rm d}{\bf x}~{\bf {j}}({\bf x})\cdot{\bf A}(t,{\bf x}),\nonumber \\
{\bf j}({\bf x})&=&\frac{e\hbar}{2 m}\left(\psi_{\alpha}({\bf x})^{\+}{i{\vec \nabla}}\psi_{\alpha}({\bf x})+h.c.\right)+{\cal O}(A),\nonumber \\
{\rm H}_{A}&=&\frac{1}{8 \pi}\int {\rm d}{\bf x} \left(E^2+B^2\right),
\label{Hpieces} 
\eer
%We find an electronic structure of the model by 
where $\psi_{\alpha}({\bf x})$ is the electron field, $\alpha,~\beta$ are the spin indices;
%%%%%
%; ${\rm \psi}$ and ${\rm \psi}^{\+}$ obey canonical 
%anticommutation relations ${\{}\psi_{\alpha}({\bf x}),\psi^{\+}_{\beta}({\bf y}){\}}=\delta_{\alpha\beta}\delta({\bf x}-{\bf y}),~{\{}\psi_{\alpha}({\bf x}),\psi_{\beta}({\bf 
%y}){\}}=0.$ 
${e,m}$ are the electron charge and mass, and ${\rm Z}_I|e|,~M_I$ are the $I^{th}$ ion charge and mass, respectively. 
$V_C$ is the bare Coulomb interaction between electrons; $\{ {\bf P}_I,{\bf R}_I \}, ~I=1,..,{\rm N}_{i},$ 
are the momentum and position operators of the ions obeying canonical commutation relations 
$[{R}_{aI},{P}_{bJ}]=i\hbar\delta_{ab}\delta_{IJ},~a,b=x,y,z,$ where ${\rm N}_{i}$ is the 
number of atoms of type $i.$ Further, ${\bf A}(t,{\bf x})$ is the electromagnetic vector potential, and 
${\bf E}(t,{\bf x}),~{\bf B}(t,{\bf x})$ are the electric and magnetic field operators, respectively. 
Coulomb gauge is implemented.
%which will be treated classically in this work, {\it i.e.}, ${\bf A}(t,{\bf x})$ is some known function; 
%${\bf j}({\bf x})$ is the current operator of which only the field independent part relevant here is explicitly shown; ${\rm c}$ is the speed of light. 
In the DFT simulations the microscopic ion-electron Coulomb potential
%, ${\rm v}({\bf x},{\bf R}_I)$, is Coulomb with ${\rm v}({\bf x},{\bf R}_I)=1/|{\bf x}-{\bf R}_I|.$ However, in practice 
%the atomic core electrons are integrated out resulting in 
is replaced by a pseudopotential ${\rm v}({\bf x},{\bf R}_I)$ felt by the valence electrons that are
explicitly included in DFT simulations ({\it e.g.}, \cite{PhysRevB.54.11169,PhysRevB.59.1758}).
%; $V_{NN}$ is also modified.
%As shown above, in the first approximation 
%So far, it has been approximated 
%for ${\rm v}({\bf x},{\bf R}_I)$ 
%and $V_{NN},$ which is the effective ion-ion interaction, we 
%one use Coulomb potential. 
%As discussed in the following, it is to be augmented by the medium screening. 

The 
%photon terms, electron-photon coupling, 
%, ${\rm H}_{A},$ 
relativistic corrections (spin-orbit coupling, fine structure, {\it etc.}) 
are not relevant to this work and have been omitted. 

\subsection{Time-domain DFT and Non-Adiabatic Molecular Dynamics}

Let us review the basic ideas of the time-domain DFT (tdDFT) approach (see, {\it e.g.}, \cite{KilinaACSNANO2009}). 
The electrons are coupled to the ions' potential which becomes time-dependent if the thermal 
motion of ions in the system is included. Time-dependent Kohn-Sham (KS) equation of time-dependent DFT 
(see, {\it e.g.}, \cite{CASIDA20093}) reads
\ber
\left(i \hbar \frac{\partial}{\partial t} + {\rm H}_{KS}({\bf x};{\bf R}(t))\right)\phi_i=0,
\label{TDKS}
\eer
where KS operator depends on time via instantaneous ion positions ${\bf R}(t).$
As a side note, in the most general case of a spin-polarized periodic system we have composite state labels
$i={\{}n,{\bf q},\sigma{\}},$ where $n$ is the band number, ${\bf q}$ is the lattice wave vector, $\sigma$ 
is the spin projection. But in this work we only consider spin-symmetric aperiodic systems where the orbitals
are labeled by just an integer. So, one seeks time-dependent KS orbitals $\phi_i$ as an 
expansion in the adiabatic KS orbitals, {\it i.e.}, in the eigenfunctions of the KS equation 
at each moment of time, so that
\ber
\left({\rm H}_{KS}({\bf x};{\bf R}(t))-\epsilon_i({\bf R}(t))\right)\varphi_i({\bf R}(t))=0.
\label{KSeq}
\eer
Then
\ber
\phi({\bf x},t)=\sum_i c_i(t) \varphi_i({\bf x};{\bf R}(t))
\label{phit}
\eer 
This leads to the equation for the coefficients $c_i$
\ber
i \hbar\frac{\partial c_i}{\partial t} = \epsilon_i c_i +  \sum_k c_k d_{ik}=0,~
d_{ik}(t)=i \hbar \int{\rm d}{\bf x}~\varphi^*_i\frac{\partial}{\partial t} \varphi_k,
\label{eqci}
\eer
where $d_{ik}(t)$ are the non-adiabatic couplings (NACs) that describe KS state coupling due to 
electron-phonon coupling; $d_{ik}$ are computed using procedures of DFT-based NAMD (see, {\it e.g.}, \cite{KilinaACSNANO2009,doi:10.1021/ct5004093}).

\subsection{Incorporating NACs into MBPT}
Let us start by assuming for a moment that the ions are classical and stationary at their equilibrium 
ion positions ${\{}{\bf R}_I^0{\}}$ corresponding to the energy minimum (found by, {\it e.g.}, the DFT 
geometry relaxation procedure). Then the hamiltonian is that of the electrons moving in the stationary 
ion backround. In reality the ions, of course, move due to the zero-point oscillations, finite temperature, 
external fields, {\it etc.} Therefore, additional terms describing ion motion and electron-ion interactions 
need to be added to the electron hamiltonian. Of course, these are the ion, ion-electron terms in the hamiltonian 
(\ref{Hmicro}). But let us contemplate an alternative respresentation of these operators that may prove 
more convenient for calculations. With this in mind, we note that application of the effective field 
theory paradigm (see, {\it e.g.}, \cite{kaplan2005lectureseffectivefieldtheory}) to this problem tells 
us that including ion motion and ion-electron interactions will generate additional local operators in 
the hamiltonian that are allowed by the symmetries of the problem. These should involve some effective 
bosonic degrees of freedom $C_{\nu}$ with $[C_{\nu},C^{\+}_{\mu}]=\delta_{\mu \nu},~[C_{\nu},C_{\mu}]=0,$ 
where $\nu$ is some label, and, also, operators that mix $C$ and electron fields. The lowest order operators 
in the effective theory describing phonons and electron-phonon interactions are
\ber
\delta H_{\nu}=\sum_{\nu} \left(\hbar \omega_{\nu} + \frac12\right) C^{\+}_{\nu} C_{\nu} + \sum_{ij\nu}\left(a_{i\sigma}^\+ a_{j\sigma}\hbar g_{ij}(\omega_{\nu})C_{\nu}+h.c.\right),
\label{Hnu}
\eer
where the energies $\hbar\omega_{\nu} \geq 0$ are continuous (or, more precisely, quasi-continuous as we will show later); 
${\rm a}_{i\alpha}$ is the annihilation operator of a fermion in the ${\rm i}^{th}$ 
KS state with spin $\alpha$ so that $\psi_{\alpha}({\bf x})=\sum_i\varphi_{i}({\bf x}){\rm a}_{i\alpha}$ \cite{AGD,Mahan}.
Further, $g_{ij}(\omega_{\nu})=g_{ji}(\omega_{\nu})^*$ are some electron-phonon coupling constants. 
Both $\omega_{\nu}$ and $g_{ij}(\omega_{\nu})$ are unspecified at this point. Of course Eq.~(\ref{Hnu}) 
is just the standard hamiltonian describing electron-phonon coupling. 
But the labels on the degrees of freedom are not associated with the normal modes of atomic oscillations. 
Operators that involve higher powers of $a_i$ and $C_{\nu}$ are suppressed by additional powers of 
$E_{thermal}/E_{electron}\sim {\rm T}/E_{g},$ where $T$ is the temperature (in the energy units) and $E_g$ is
the electron energy gap. In the practically relevant case of sufficiently low temperature and finite gap 
${\rm T}/E_{g} \ll 1$ and it is sufficient to use Eq.~(\ref{Hnu}). 

Let us now relate couplings $g_{ij}(\omega_{\nu})$ and the NACs of NAMD. For the moment let us treat the effects of
nuclear motion as an external time-dependent potential coupled to the electrons. The lowest order 
operator describing the coupling of electrons to a time-dependent potential ${\rm V}({\bf x},t)$ is
\ber
\delta H_{NAC}=\int_t a_{i\sigma}^\+ v_{ij}(t) a_{j\sigma}, \text{where}~v_{ij}(t)=\int {\rm d}{\bf x}\varphi_i^*({\bf x}) {\rm V}({\bf x},t)\varphi_j({\bf x}).
\label{dHNAC}
\eer
Note that in the above, and, more generally, in MBPT calculations there is some ambiguity in the  
choice of KS basis orbitals. In principle, one may use instantaneous adiabatic KS orbitals 
${\{}\varphi_{i}({\bf x};{\bf R}_I(t)){\}}$ at any $t.$ In the case of thermal oscillations of 
the ions about well-defined equilibrium positions $\varphi_{i}({\bf x};{\bf R}^0_I)$ is a reasonable choice. 
If significant structural rearrangements, accompanied by significant electronic structure changes, such as 
bond breaking, occur it may be necessary to update $\varphi$ following each such event. This may result in 
having to re-solve $G_0W_0$ and BSE in the basis of updated orbitals. In this work we will be assuming 
the simplest case of small thermal oscillations and, so, we use ${\{}\varphi_{i}({\bf x};{\bf R}^0_I){\}}.$ 
Let $\varphi_{i}({\bf x};{\bf R}^0_I) \equiv \varphi_{i}({\bf x}).$ 

%\ber
%\psi_{\alpha}({\bf x})=\sum_i\phi_{i\alpha}({\bf x}){\rm a}_{i\alpha}.
%\label{psi_to_a}
%\eer
Now let us note that $v_{ij}(t)$ is the amplitude for $\ket{j}\to\ket{i}$ phonon-mediated transition which 
in tdDFT NAMD is given by the NAC. This is what allows us to identify $v_{ij}$ as $d_{ij}:~v_{ij}(t) \equiv d_{ij}(t).$ 

Our main goal is to compute contribution to the phonon-mediated relaxation collision integrals in BE that include NACs. 
For that we need $\Sigma^{+-},~\Sigma^{-+},$ which are the self-energies of the Keldysh technique \cite{Landau10,doi:10.1063/1.4997048}. 

It is instructive to start by computing the zero-temperature contribution to the electron self-energy due to NACs treated as 
an external time-dependent potential as in Eq. (\ref{dHNAC}) and then to generalize to the finite-temperature nonequilibrium case. 
The issue that arises here is that inclusion of $d_{ij}(t)$ apparently violates energy conservation and, therefore, time translation 
invariance that should be present in the correlation functions that describe all the electron-phonon processes. 
So, some additional manipulations will be needed to define correct NAC contribution to the electron self-energy. Once the 
NAC-dependent electron self-energy expression is obtained, we will deduce the effective hamiltonian for electron-phonon coupling 
by comparison with the effective hamiltonian (\ref{Hnu}). Then the approach will be generalized to the finite-temperature nonequilibrium case
leading to the expressions for the collision integrals. 

So, the leading order NAC contribution to the KS state propagator $G(t_2,t_1)=\bra{g.s.}{\rm T}a_i(t_2)a^{\+}_i(t_1)\ket{g.s.}$ comes 
from the ${\cal O}\left(d_{ij}^2\right)$ correction. It reads 
\ber
\delta G(\omega_2,-\omega_1)=g_i(\omega_2) {\cal T}^2\sum_k \int {\rm d}\omega^{'} d_{ik}(\omega_2-\omega^{'})g_k(\omega^{'})d_{ki}(\omega^{'}-\omega_1)g_i(\omega_1),
\label{dGit2t1}
\eer
where $g_k(\omega)=i/(\omega-[\epsilon_k-\mu]+i\delta_k)$ is the KS state propagator, $\mu$ is the grand-canonical chemical potential (set at mid-gap), $\delta_k=\delta~{\rm sgn}\left(\epsilon_k-\mu\right)$ \cite{AGD};
\ber
d_{ij}(\omega)=\frac{1}{{\cal T}}\int_0^{\cal T} {\rm d}t e^{ i \omega t} d_{ij}(t),
\label{dijw}
\eer
where ${\cal T}$ is the time span over which NACs are computed in NAMD. For NACs we assume ``periodic boundary conditions" $d_{ij}(t)=d_{ij}(t+{\cal T}),$ which 
is a benign assumption given that ${\cal T}$ is much longer that any intrinsic time scale. 
Next, we recall that all the elementary electron-phonon processes conserve energy, which implies that the 
electron correlator should only depend on the part of $d_{ik}(t_1)d_{ki}(t_2)$ that respects the time-translation invariance, {\it i.e.}, 
is a function of $t_1-t_2.$ These considerations result in the following electron self-energy
\ber
\Sigma_{NAC}(\omega)=-i {\cal T} \sum_k \int {\rm d}\omega^{'} |d_{ik}(\omega^{'})|^2 g_k(\omega-\omega^{'}).
\label{SigmaNAC}
\eer  
We note that the product of NAC potentials in the integrand is related to the Fourier transform of the NAC autocorrelator
used in the reduced density matrix Redfield theory approach \cite{doi:10.1021/ct5004093}.
%mention ergodicity
Next, $\Sigma_{NAC}$ can be rewritten as
\ber
\Sigma_{NAC}(\omega)= - \sum_k \int \frac{{\rm d}\omega^{'}}{2 \pi}\frac{{\rm d}\omega_{\nu}}{2 \pi i} {\cal T} |d_{ik}(\omega_{\nu})|^2 g_k(\omega-\omega^{'})D_{{\nu}}(\omega^{'}),             
\label{SigmaNACOmega}
\eer
where $D_{{\nu}}(\omega)=i/(\omega-\omega_{\nu}+i\delta)$ is a propagator of a bosonic state with energy $\hbar \omega_{\nu}.$ This expression could have come as the 
${\cal O}\left(d_{ij}^2\right)$ correction correction from the electron Hamiltonian augmented with 
\ber
\delta H_{\nu}=\sum_{\nu} \left(\hbar \omega_{\nu} + \frac12\right) C^{\+}_{\nu} C_{\nu} + \sum_{ij\nu}\left(a_{i\sigma}^\+ a_{j\sigma}\hbar d_{ij}(\omega_{\nu})C_{\nu}+h.c.\right),
\label{Hnu1}
\eer
where $C_{\nu}$ are some bosonic operators; the frequencies $\omega_{\nu}$ can be treated as quasi-continuous
with $$\frac{\cal T}{2 \pi}\int {\rm d}\omega_{\nu} = \sum_{\nu}.$$
The operators in Eq.~(\ref{Hnu1}) are precisely the leading order operators in the effective 
theory describing electron-phonon interactions shown in Eq.~(\ref{Hnu}). 
%The manipulations leading to Eq.~(\ref{Hnu1}) have 
This allows us to identify $d_{ij}(\omega_{\nu})$ as the effective coupling constants $g_{ij}(\omega_{\nu})$ in Eq.~(\ref{Hnu}): $g_{ij}(\omega_{\nu})\equiv d_{ij}(\omega_{\nu}).$ 
The higher order operators involving more powers of $a_i$ and $C_{\nu}$ are suppressed by additional powers of $E_{thermal}/E_{electron}\sim {T}/E_{g}.$ 
In the practically relevant case of low temperature and finite electron spectrum gap ${T}/E_{g} \ll 1$ and it is sufficient to
use Eq.~(\ref{Hnu1}). 

The last step is to include the electron-hole bound states. It is known that in order to include effects of excitons the hamiltonian 
(\ref{Hmicro}) needs to be augmented with the following exciton and exciton-electron terms \cite{SFJCP}
%See Appendix B for definitions.
\ber
\delta{\rm H}_{e-exciton}&=&
\sum_{e h \alpha}\sum_{\sigma} \frac{1}{\sqrt{2}}\left(\left[\epsilon_{eh}-E^{\alpha}\right]{\rm \psi}^{\alpha}_{eh}a_{h\sigma}a^{\dagger}_{e\sigma}
({\rm B}^{\alpha}+{\rm B}^{\alpha\dagger})+h.c.\right)+\nonumber \\&+&\sum_{\alpha}E^{\alpha}{\rm B}^{\alpha\dagger}{\rm B}^{\alpha},~\epsilon_{eh}=\epsilon_{e}-\epsilon_{h},
\label{Heexc}
\eer
where ${\rm B}^{\alpha\dagger},~E^{\alpha}$ are the singlet exciton~creation~operators and energies, respectively; a spin-zero exciton state is represented as 
%can be expanded in the basis of electron-hole states as 
\cite{PhysRevB.62.4927,PhysRevB.29.5718}
\ber
\ket{{\alpha}}_{0}={\rm B}^{\alpha\dagger}\ket{g.s.}=\sum_{e h}\sum_{\sigma=\uparrow,\downarrow}\frac{1}{\sqrt{2}} {\rm \psi}^{\alpha}_{eh} a^{\dagger}_{e\sigma} a_{h\sigma} \ket{g.s.},
%~\sigma=\uparrow,\downarrow,
\label{Psialpha}
\eer 
where ${\rm \psi}^{\alpha}_{eh}$ 
%and $E^{\alpha}$ are 
is the spin-zero exciton wavefunction; the index ranges are $e>HO,~h\leq HO,$ where HO 
%and LU are 
is the highest occupied
% and lowest unoccupied 
KS level, $LU=HO+1$ is the lowest unoccupied KS level.
To determine
%include dynamics of \cite{
exciton wave functions and energies one solves the Bethe-Salpeter equation (BSE) in the Tamm-Dancoff approximation and using static approximation 
for the screened interaction \cite{PhysRevB.62.4927,PhysRevB.29.5718}. It is understood that one should avoid double counting when using ``mixed" 
electron-exciton Hamiltonian \cite{SFJCP}. Here we only include singlet exciton states. Inclusion of spin-1 exciton states is straightforward \cite{SFJCP}.

Having identified the hamiltonian terms responsible for the electron-phonon coupling, it is a standard exercise to extend the setting to the 
weak nonequilibrium at finite-temperatue and to compute the corresponding Boltzmann transport equation collision integrals using Keldysh 
technique self-energies (see, {\it e.g.}, \cite{doi:10.1063/1.4997048}). The expressions will be presented in the following subsection.

Let us discuss the main simplifying approximation made in this work. In our DFT simulations we have used Vienna Ab-Initio Simulation Package (VASP) software \cite{PhysRevB.54.11169} 
with the hybrid Heyd-Scuseria-Ernzerhof (HSE06) exchange correlation functional \cite{vydrov:074106,heyd:219906}, 
which has been somewhat successful in reproducing electronic gaps in various semiconductor nanostructures ({\it e.g.}, 
\cite{RevModPhys.80.3,PhysRevLett.107.216806}.) So, here using the HSE06 functional is to 
substitute for computing $G_0W_0$ corrections to the KS energies, {\it i.e.}, for the first step in the standard three-step 
procedure \cite{PhysRevB.34.5390,PhysRevB.62.4927}. Therefore, 
%here we use the simplest approximation where 
single-particle energy levels and orbitals are approximated by the KS $\epsilon_i$ and 
$\phi_i({\bf x})$ from the HSE06 DFT output. 
%{\color{red} This implies that fermion lines in the Feynman diagrams are treated as ``dressed'', 
%{\it i.e.}, the self-energy corrections as well as the compensating 
%${\rm H}_V$ term (\ref{HV}) have been incorporated in the Fermion propagators. \cite{RevModPhys.74.601}.} 
The rational for this major simplifying approximation is that the emphasis of this work is on the method development. 
Also, while inclusion of $G_0W_0$ technique would improve accuracy of our calculations, it is unlikely to alter our results and conclusions qualitatively. 
It is straightforward to incorporate $G_0W_0$ into this approach, which will be done at a later stage of this effort.
%{The approach and approximations used in this work}

NACs for the systems used in this work have been collected from the NAMD runs at $T=300~K$ using previously 
developed methods. Specifically, we have been using software developed by the group of Prof.~Kilin (NDSU), which is based on VASP,
and using the Perdew-Burke-Ernzerhof (PBE) exchange-correlation functional \cite{PhysRevB.54.11169,PhysRevLett.77.3865}.

\subsection{Expressions for the Collision Integrals}
Dynamics of a photoexcited nanoparticle is determined by the interplay of the phonon-mediated transitions 
between low-energy exciton states, exciton transfer and recombination, both radiative and non-radiative. It can 
be described by solving BE for the time evolution of the occupancies of the low-energy electronic states with the 
collision integrals that include all the relevant time-evolution channels. 
This results in the system of equations for the occupation numbers, which is
\ber
\frac{d n_{\alpha}}{d t}&=&{\cal I}_{\text{phonon}}[ n_{\alpha}]+{\cal I}_{\text{exciton~transfer}}[ n_{\alpha}]+{\cal I}_{\text{photon}}[ n_{\alpha}], 
\label{BE}
\eer
where $n_{\alpha}(t)$ is the occupation number of the exciton state $\alpha$ with energy $E_{\alpha}=\hbar\omega_{\alpha};$ 
${\cal I}$s in the right-hand side are the contributions to the collision integrals from different competing time-evolution channels. 
Here we include phonon-mediated relaxation, energy (or exciton) transfer and photon-mediated processes, such as 
photon recombination and photon-mediated exciton transitions. Other processes, such as carrier multiplication, are not included 
here and left to future work.  

The phonon terms are
\ber
{\cal I}_{\text{phonon}}&=&\sum_{\beta}R_{\alpha \beta }(\omega_{\alpha \beta}) \left(n_{\alpha\beta } n_{\beta } - n_{\alpha } \left(1+n_{\beta }+n_{\alpha\beta}\right)\right)
\theta(\omega_{\alpha \beta}) \cr &&
+ \sum_{\beta}R_{\alpha \beta }(\omega_{\beta \alpha}) \left(n_{\beta} \left(1+n_{\alpha}+n_{\beta\alpha}\right) - n_{\alpha}n_{\beta\alpha}\right)
\theta(\omega_{\beta \alpha}) + {\cal I}_{\text{nonrad.}}, \nonumber \\ && \omega_{\alpha\beta}=\omega_{\alpha}-\omega_{\beta},
\label{Rphonon}
\eer
where $R_{\alpha \beta}$ are the NAC couplings expressed in the exciton basis which are defined below in Eq.~(\ref{Galphabeta}); $\theta(x)$ 
is the step function. Here $n_{\alpha\beta}$ are the occupation numbers of the effective bosonic states with energy 
$\hbar \omega_{\nu}=\hbar \omega_{\alpha\beta}$ introduced in Eq.~(\ref{Hnu}). Strictly speaking, BE should be augmented with the 
terms describing time evolution of the effective boson states occupation numbers  
%with occupation numbers $n_{\nu}=<C^{\+}_{\nu} C_{\nu}>$ 
due to their interactions with the electrons. But here we will just use the unperturbed equilibrium values
$$n_{\alpha\beta}=\left({\rm exp}\left(\frac{\hbar  \omega_{\alpha\beta}}{T}\right)-1\right)^{-1}$$
since the corrections to the equilibrium $n_{\alpha\beta}$ are ${\cal O}\left({T}/E_{g}\right).$
The couplings describing exciton-exciton phonon-mediated transitions, that is the NACs in the exciton basis, are
\ber
R_{\alpha\beta}&=&
%|
\abs*{D_{\alpha\beta}-F_{\alpha\beta}}^2 {\cal T},\cr 
D_{\alpha\beta}&=&\sum_{e h_1 h_2}d(\omega_{\alpha\beta})_{h_1 h_2} \left(\psi_{e{h_1}}^{\alpha}\right){}^* \psi_{e{h_2}}^{\beta} 
+ d(\omega_{\alpha\beta})_{{e_1 e_2}} \left(\psi_{{e_1}h}^{\alpha }\right){}^* \psi_{{e_2}h}^{\beta},
\cr
F_{\alpha\beta}&=&\sum_{e h_1 h_2}\frac{d(\omega_{\alpha\beta})_{{h_1 h_2}} \psi_{e{h_1}}^{\alpha} \left(\psi_{e{h_2}}^{\beta }\right){}^*
\left(\omega_{e{h_1}}-\omega_{\alpha}\right)
\left(\omega_{e{h_2}}-\omega_{\beta}\right)}{\left(\omega_{e{h_1}}+\omega_{\alpha}\right)
\left(\omega_{e{h_2}}+\omega_{\beta }\right)}+ \cr &+&
\sum_{e_1 e_2 h}\frac{d(\omega_{\alpha\beta})_{{e_1 e_2}} 
    \psi_{{e_1} h}^{\alpha } \left(\psi_{{e_2} h}^{\beta}\right){}^* \left(\omega_{{e_1} h}-\omega_{\alpha}\right) 
   \left(\omega_{{e_2} h}-\omega_{\beta}\right)}{\left(\omega_{{e_1} h}+\omega_{\alpha}\right) \left(\omega_{{e_2} h}+\omega_{\beta}\right)},
\nonumber \\ &&\omega_{i j}=\left(\epsilon_{i}-\epsilon_{j}\right)/\hbar, 
\label{Galphabeta}
\eer
where all NAC couplings are evaluated on the corresponding exciton energy difference.
%: $d_{i j} \equiv d_{i j}(\omega_{\alpha\beta}).$
Note that the leading terms in Eq. (\ref{Galphabeta}) have the expected ``chain-rule" structure that follows from resolving 
$\bra{\alpha} \frac{d}{dt} \ket{\beta}$ in terms of NACs for electron and hole states \cite{jiang2021real}.

The last term in (\ref{Rphonon}) is supposed to include non-radiative or ``dark" exciton recombination, 
which needs to be included in order to accurately predict observables, such as PL and quantum yield. However, 
the main objective of this work is the method development, which is the rationale for why inclusion of ${\cal I}_{\text{nonrad.}}$ is left
to future work.

Expressions for the exciton transfer contribution ${\cal I}_{\text{exciton~transfer}}$ have been previously introduced elsewhere \cite{doi:10.1021/acs.jpclett.8b02288}.
The terms in ${\cal I}_{\text{photon}}$ responsible for the exciton-to-photon recombination and 
photon-mediated exciton-to-exciton transitions - $\alpha \to \beta +\gamma$ - are
\ber
{\cal I}_{\text{photon}}&=&-n_{\alpha } R^{{recomb}}_{\alpha }-\sum_{\beta }
   \theta (\omega_{\alpha \beta }) n_{\alpha } \left(1+n_{\beta}\right) R_{\alpha \to \beta +\gamma}+\\ \nonumber &+& \nonumber \sum _{\beta }
   \theta(\omega_{\beta \alpha }) \left(1+n_{\alpha}\right)
   n_{\beta } R_{\beta \to \alpha +\gamma},\\ \nonumber
   {\bf J}_{ij}&=&\sum_{{\bf p}}\varphi_i^*({\bf p}){\bf p}\varphi_j({\bf p}).
\label{BErad}
\eer
The recombination rate for the exciton-to-photon process is
\ber
R^{recomb}_{\beta}=\frac{4 \hbar ^2 \omega _{\beta } \alpha _{{fs}}}{3 c^2 m^2} \sum_{e h}\sum_{a=x,y,z}\abs*{J^a{}_{{eh}} \left(\psi_{{eh}}^{\beta }\right){}^*}^2,
\label{Rrad}
\eer
where $\alpha_{{fs}}\simeq 1/137$ is the fine structure constant.
%%%%%%%%%%%%%%%%%%%%%%%%%%%%%%%%%%%%%%%%%%%%%%%%%%%
%\raisebox{0.175\totalheight}{\includegraphics*[width=0.45\textwidth] {trion.eps}}

\begin{figure}[!t]
\vspace{-0.5cm}
\center

\begin{tabular}{cc}

$Pb_{68}Se_{68}$ & $Cd_{37}Pb_{31}Se_{68}$\\

\raisebox{0.175\totalheight}{\includegraphics*[width=0.475\textwidth] {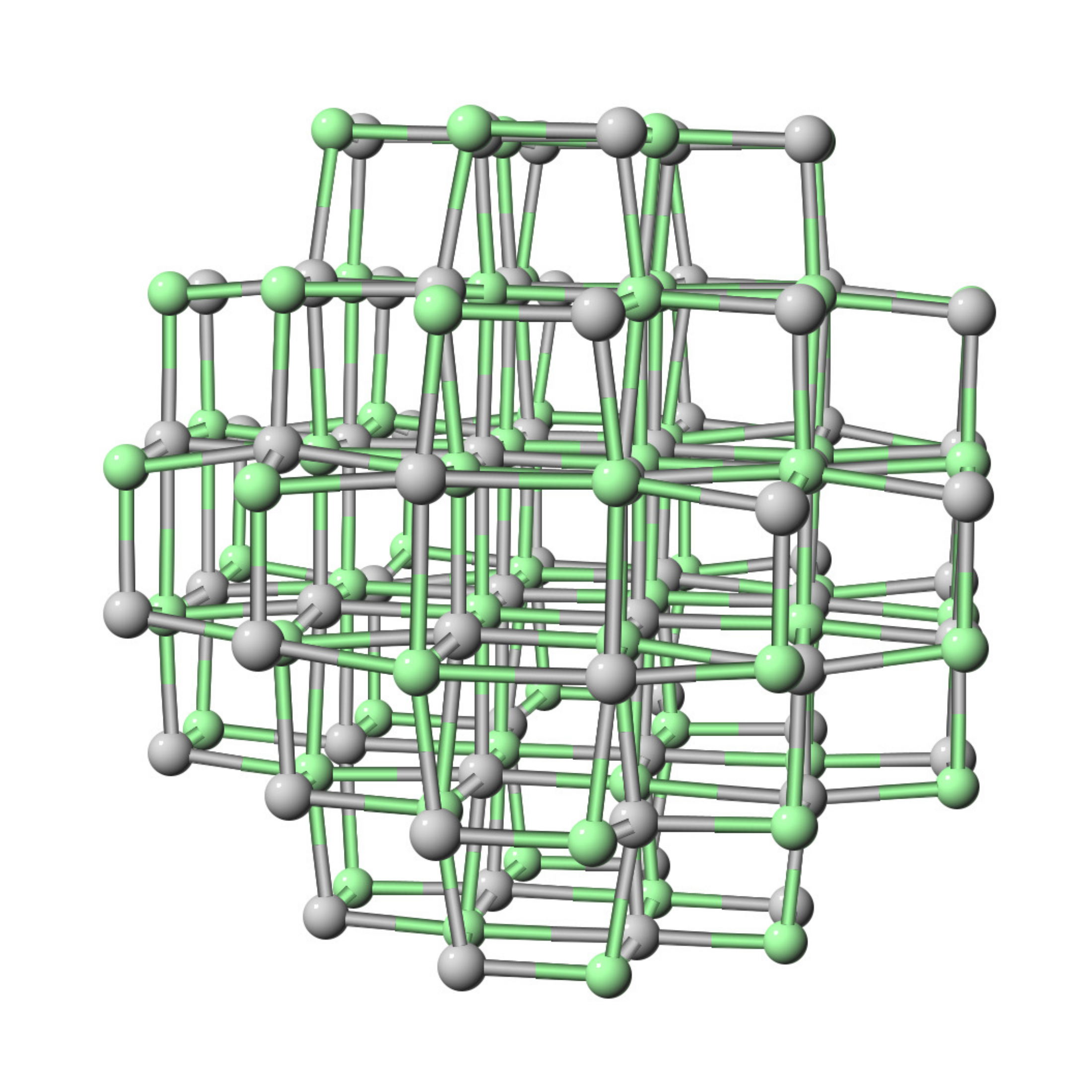}} &

%\hfill

\raisebox{0.175\totalheight}{\includegraphics*[width=0.485\textwidth] {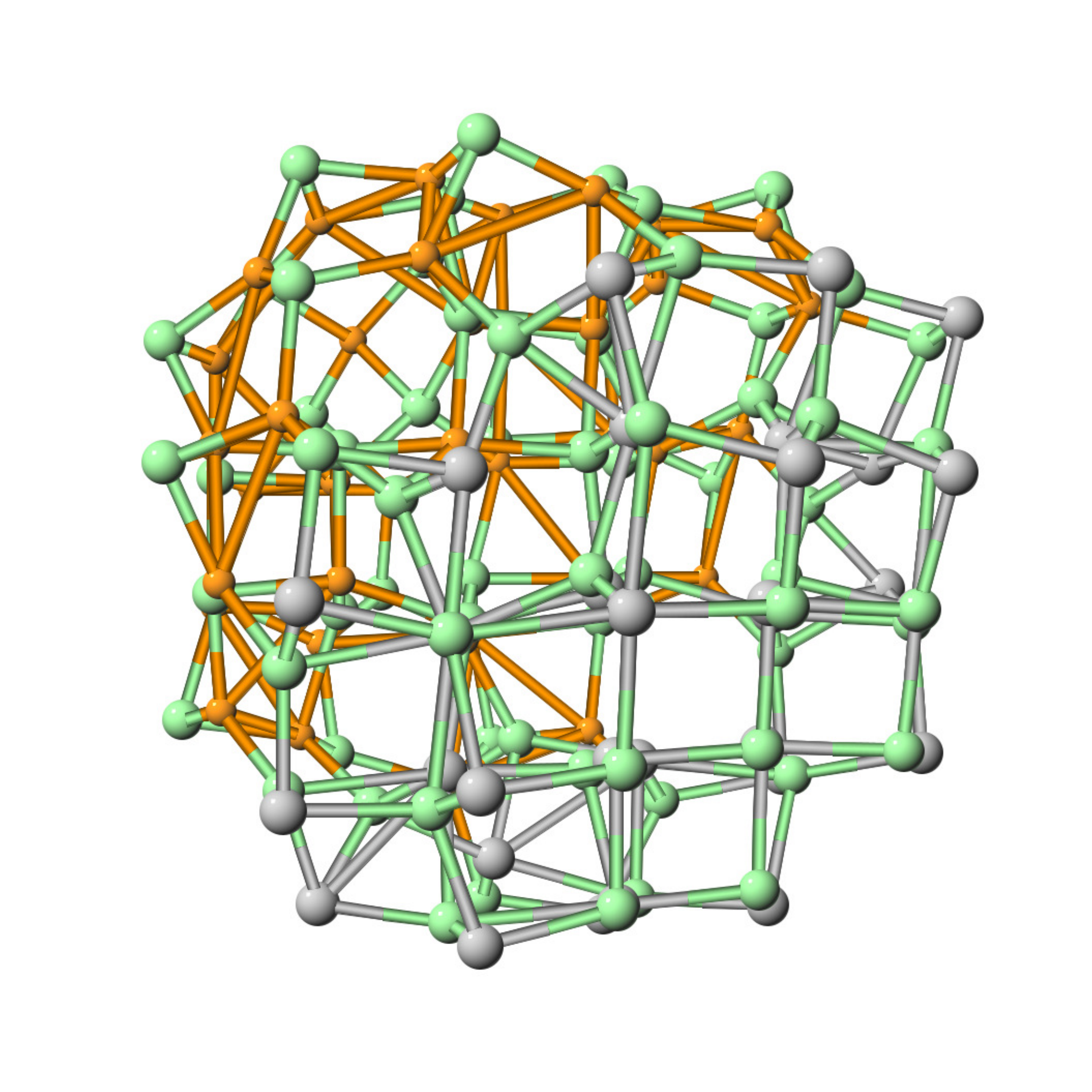}}\\

\end{tabular}

\vspace{-1.5cm}

\caption{ 
Atomistic models for $Cd_{37}Pb_{31}Se_{68}$ and $Pb_{68}Se_{68}$.  
}

\vspace{-2.25ex}

\label{fig:Cd37_Pb68}

\end{figure}

%%%%%%%%%%%%%%%%%%%%%%%%%%%%%%%%%%%%%%%%%%%%%%%%%%%
As mentioned above, another process included here is the photon-mediated exciton-to-exciton transition where a soft photon is emitted, which is the 
leading order contribution to the exciton bremsstrahlung. The expression for the rate
$R_{\alpha \to \beta +\gamma}$ is
%%%%%%%%%%%%%%%%%%%%%%%%%%%%%%%%%%%%%%%%%%%%%%%%%%%
%\raisebox{0.175\totalheight}{\includegraphics*[width=0.45\textwidth] {trion.eps}}

\begin{figure}[!t]
\vspace{-0.001cm}
\center

\begin{tabular}{cc}

$Pb_{68}Se_{68}$ & $Cd_{37}Pb_{31}Se_{68}$\\

\raisebox{0.175\totalheight}{\includegraphics*[width=0.475\textwidth] {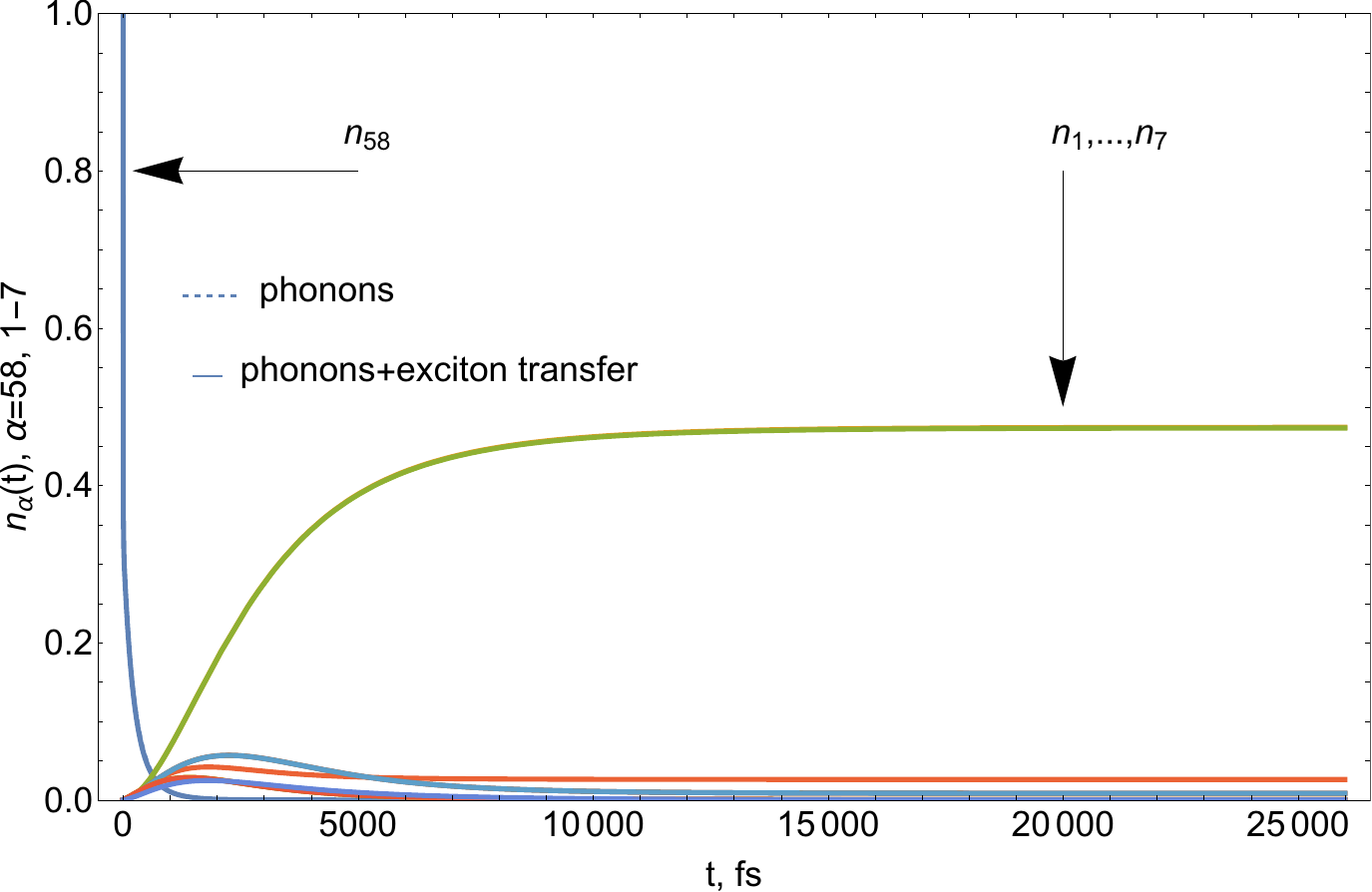}} &

%\hfill

\raisebox{0.175\totalheight}{\includegraphics*[width=0.485\textwidth] {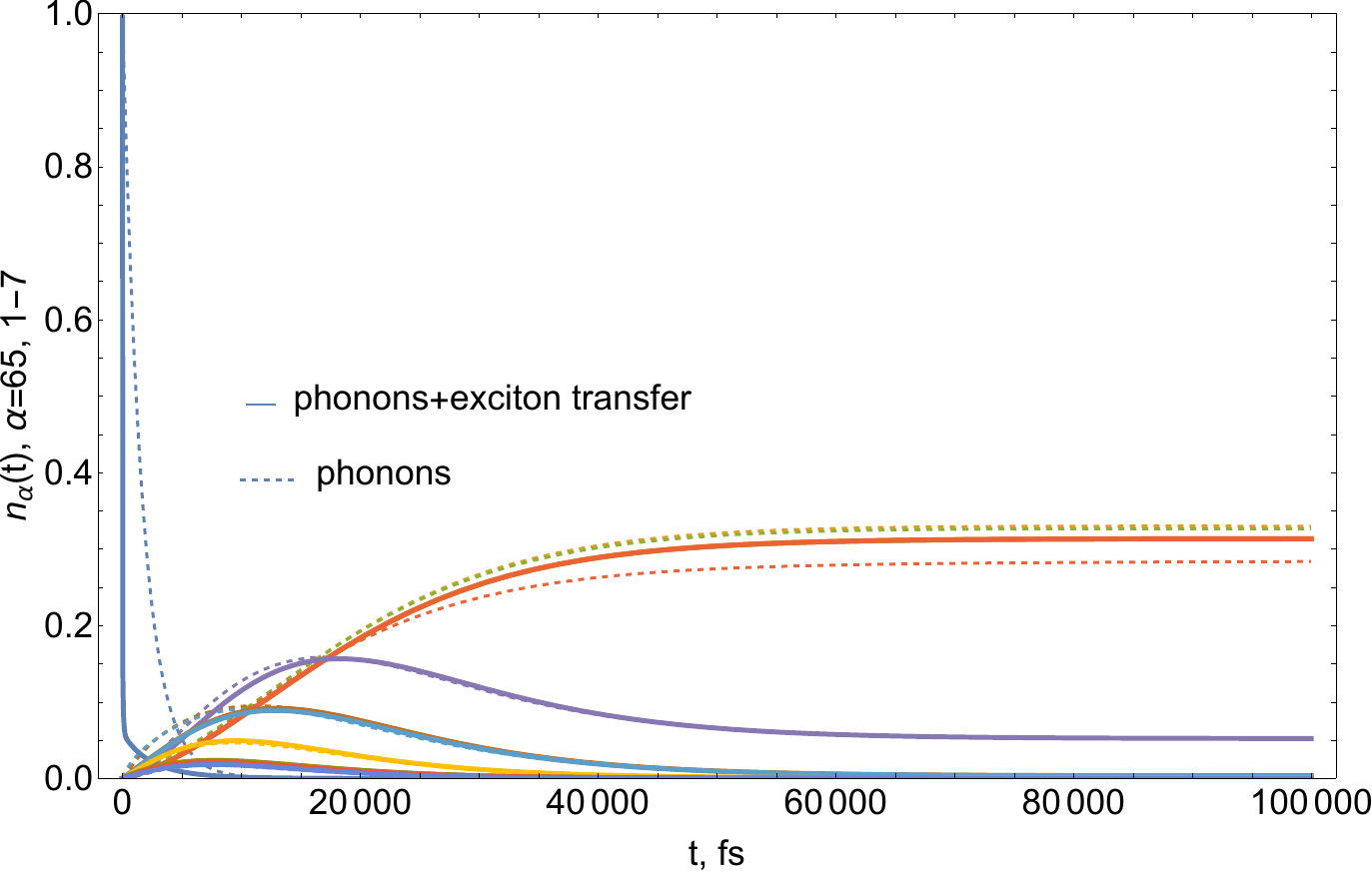}}\\

\end{tabular}

\vspace{-1.0cm}

\caption{ 
Low-energy exciton occupation numbers for Janus NC $Cd_{37}Pb_{31}Se_{68}$ and $Pb_{68}Se_{68}$.  
}

\vspace{-2.25ex}

\label{fig:n_Cd37_Pb68}

\end{figure}

%%%%%%%%%%%%%%%%%%%%%%%%%%%%%%%%%%%%%%%%%%%%%%%%%%%
\ber
R_{\alpha \to \beta +\gamma}&=&\frac{4 \hbar ^2 \alpha_{\text{fs}}\omega_{\alpha\beta}}{3 c^2 m^2} \sum_{e h}\sum_{a=x,y,z}\abs*{{\cal J}_{\alpha\beta}^a+{\cal L}_{\alpha\beta}^a}^2,\nonumber \\ 
{\cal J}_{\alpha\beta}^a&=&J^a{}_{e_1 e_2} \left(\psi _{e_1 h}^{\alpha }\right){}^* \psi _{e_2
   h}^{\beta }-J^a{}_{h_2 h_1} \left(\psi _{e h_1}^{\alpha
   }\right){}^* \psi _{e h_2}^{\beta },\nonumber \\ 
{\cal L}_{\alpha\beta}^a&=&\frac{J^a{}_{e_1 e_2} \psi ^{\alpha }{}_{e_1 h} \left(\psi _{e_2 h}^{\beta}\right){}^* \left(\omega _{e_1
   h}-\omega _{\alpha }\right)  \left(\omega _{e_2 h}-\omega _{\beta
   }\right)}{\left(\omega _{\alpha }+\omega _{e_1 h}\right)
   \left(\omega_{\beta }+\omega_{e_2 h}\right)}- \nonumber \\ &-&
   \frac{J^a{}_{h_1 h_2} \psi_{e {h}_1}^{\alpha } \left(\psi _{e h_2}^{\beta}\right){}^*
   \left(\omega_{e{h}_1}-\omega _{\alpha }\right) 
   \left(\omega_{e{h}_2}-\omega _{\beta }\right)} {\left(\omega_{\alpha}+\omega
   _{e{h}_1}\right) \left(\omega _{\beta }+\omega
   _{e{h}_2}\right)}.
\label{Rabrad}
\eer
%%%%%%%%%%%%%%%%%%%%%%%%%%%%%%%%%%%%%%%%%%%%%%%%%%%
\begin{figure}[!t]%!t
\vspace{-0.5cm}
\center
\raisebox{0.185\totalheight}{\includegraphics*[width=0.693965\textwidth]
{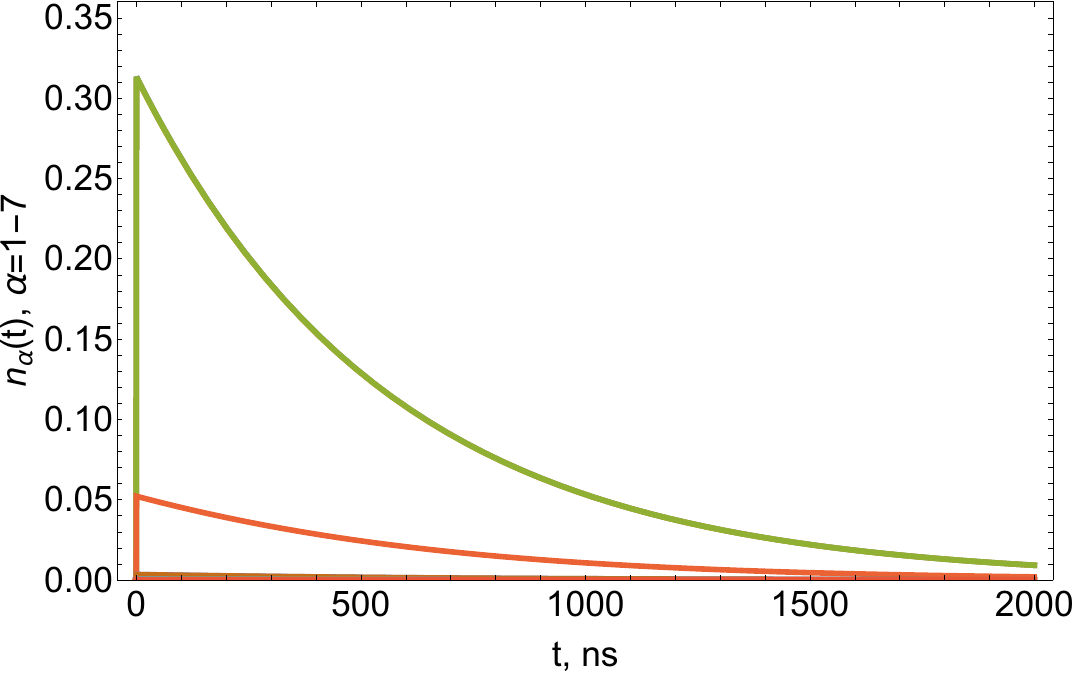}}
\vspace{-1.75cm}
\caption{
Low-energy exciton occupation numbers for Janus NC $Cd_{37}Pb_{31}Se_{68}$ including radiative recombination. 
}
%\vspace{-0.25ex}
\label{fig:nrecomb}
\end{figure}
%%%%%%%%%%%%%%%%%%%%%%%%%%%%%%%%%%%%%%%%%%%%%%%%%%%
\section{Photoluminescence Spectrum from Boltzmann Transport Equation}
First we note that \\
(1) $R^{recomb}_{\alpha} n_{\alpha}(t)$ is the contribution to the rate of decay of  
exciton $\ket{\alpha}$ due to $\ket{\alpha}\to \text{photon}$ process, {\it i.e.}, the ground state recombination; 
the emitted photon energy is $E_{\alpha}$; \\
(2) $R_{\alpha\beta}(1+n_{\beta}(t))n_{\alpha}(t)$ is the contribution to the rate of decay of exciton $\alpha$ into exciton 
 $\beta$ due to $\ket{\alpha}\to \ket{\beta} + \text{photon}$ process; the photon energy is $E_{\alpha}-E_{\beta}$. \\
Therefore, in order to compute $PL(\omega)$ one solves BE for the exciton occupancies $n_{\alpha}(t)$ using the complete collision integrals.
%, {\it i.e.}, phonon-mediated relaxation,exciton transfer, carrier multiplication, radiative and non-radiative recombination, radiative transitions, {\it etc.} 
To get $PL(\omega)$ one evaluates the terms above on the solutions $n_{\alpha}$, averages over 
the time interval over which radiative decay occurs, bins the contributions according to the emitted photon energy
$\hbar \omega$ and adds them up. After normalization this represents the fraction of emitted photon energy as a function of $\hbar \omega$. This 
generalizes the approach of \cite{doi:10.1021/jz400760h}.
%%%%%%%%%%%%%%%%%%%%%%%%%%%%%%%%%%%%%%%%%%%%%%%%%%%
%\raisebox{0.175\totalheight}{\includegraphics*[width=0.45\textwidth] {trion.eps}}
%\vspace{-2.25ex}
\begin{figure}[!t]%!t
\center
\raisebox{0.185\totalheight}{\includegraphics*[width=0.693965\textwidth] 
{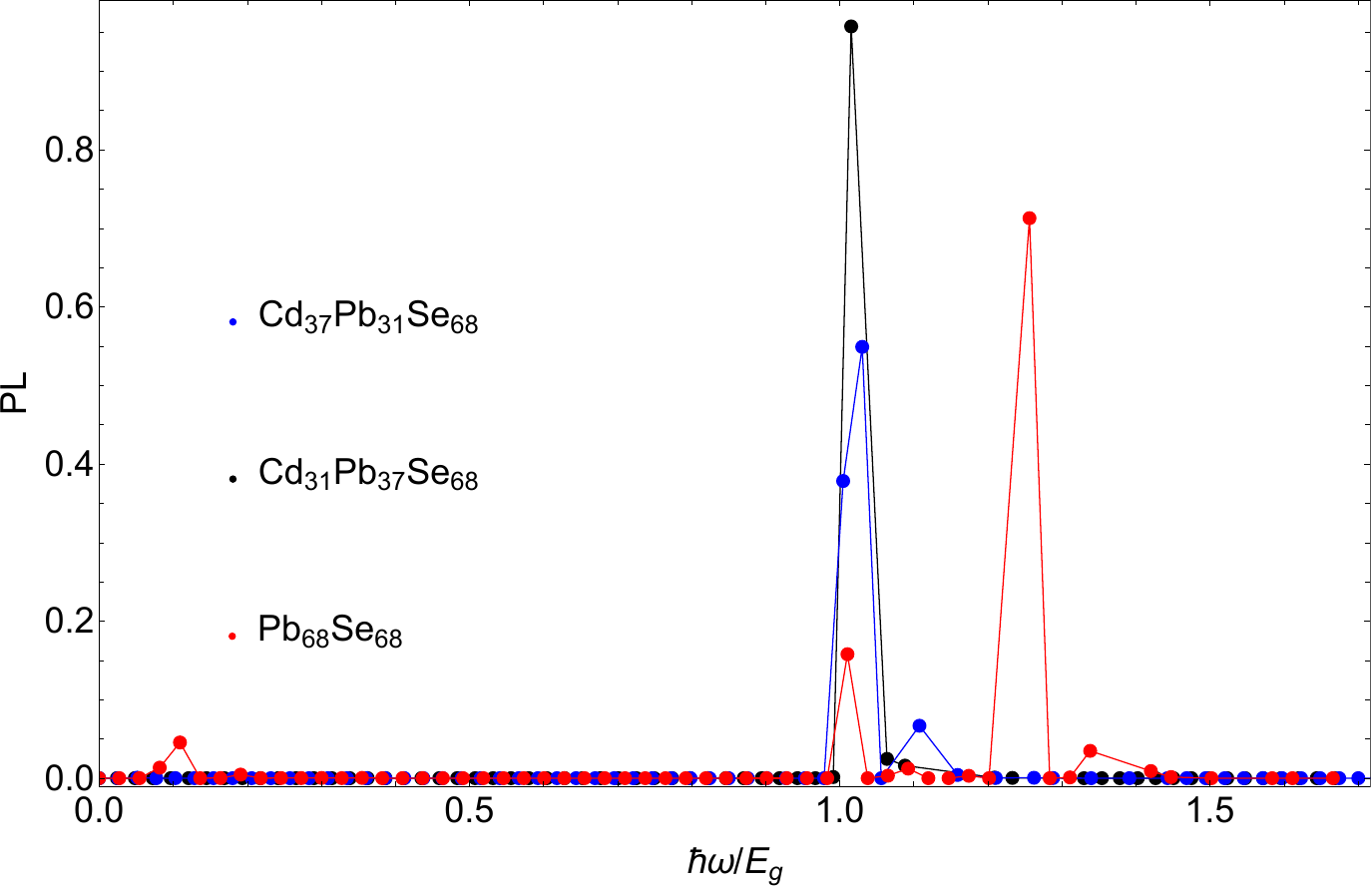}}
\vspace{-1cm}
\caption{ 
PL spectra for Janus NC $Cd_{37}Pb_{31}Se_{68},~Cd_{31}Pb_{37}Se_{68}$ and $Pb_{68}Se_{68}$. 
}
%\vspace{-0.25ex}
\label{fig:pl}
\end{figure}
%%%%%%%%%%%%%%%%%%%%%%%%%%%%%%%%%%%%%%%%%%%%%%%%%%%

\section{Results}
Here we apply the techniques developed above to make some predictions for excited state properties in 1.5 $nm$ chalcogenide nanoparticles, 
such as Janus NC $Cd_{37}Pb_{31}Se_{68},~Cd_{31}Pb_{37}Se_{68}$ and $Pb_{68}Se_{68},$ which is a reference homogeneous system of a similar size. 
Atomistic models of $Cd_{37}Pb_{31}Se_{68}$ and $Pb_{68}Se_{68}$ are shown in Fig.~\ref{fig:Cd37_Pb68}.
The BSE predictions for the exciton gaps for the three systems are $E_g=0.97~eV,1.03~eV,~0.92~eV,$ respectively.
Shown in Fig.~\ref{fig:n_Cd37_Pb68} are the predictions for the time evolution of occupation numbers of low-energy excitons on a $\sim 100~ps$ time 
scale where radiative recombination can be neglected. The results for $Pb_{68}Se_{68}$ are shown on the left, while $Cd_{37}Pb_{31}Se_{68}$ - on the right of Fig.~\ref{fig:n_Cd37_Pb68}. 
The results for $Cd_{31}Pb_{37}Se_{68}$ and $Cd_{37}Pb_{31}Se_{68}$ are similar.
The initial occupied exciton corresponds to a prominent low-energy absorption peak; then the system loses energy via phonon emission and goes 
through multiple short-lived transient states, which are not shown. Eventually, it comes to a quasi-steady state where few low-energy exciton states are occupied.
We note that for the heterogeneous Janus NCs inclusion of exciton transfer leads to notable quantitative differences in the quasi-steady state.
Also, we note for the Janus NCs relaxation, {\it i.e.}, the onset of the steady state, is significantly slower compared to that of the homogeneous NC.
This is as expected since disorder tends to decrease efficiency of electron-phonon coupling.

Shown in Fig.~\ref{fig:nrecomb} are the predictions for the time evolution of occupation numbers of low-energy excitons in $Cd_{37}Pb_{31}Se_{68},$ which 
is a representative case, on a $\sim 1000~ns$ time scale where radiative recombination decay is taking place. The quasi-stationary state formed in 
the course of phonon-mediated relaxation where few low-energy excitons were occupied is decaying exponentially due to emission of radiation. 
 
Shown in Fig. \ref{fig:pl} are the predictions for the $PL(\omega)$ of Janus NC $Cd_{37}Pb_{31}Se_{68},~Cd_{31}Pb_{37}Se_{68}$ and $Pb_{68}Se_{68}$ computed 
applying the procedure outlined above with the initial state corresponding to a low-energy absorption peak.
The emission is predicted to occur mostly at the bandgap for the Janus NCs, which is in qualitatative agreement with the available
experimental data for 5-$nm$ Janus NCs \cite{doi:10.1021/acsnano.5b01859}. For $Pb_{68}Se_{68}$ the strongest emission is predicted to occur at about 
$1.2 E_g.$ For all three systems there is also non-zero infrared emission at about $0.1 E_g.$

\section{Conclusions and Outlook}
We have argued that NACs of DFT-based NA MD can be incorporated into the DFT-based 
Keldysh approach thus resulting in the BE collision integral for phonon-mediated
carrier relaxation. This technique does not rely on perturbation theory in
coupling of electrons to individual phonon modes and harmonic approximation thus 
promising applicability to a wider class of (nano)materials. The resulting BE that 
included phonon-mediated relaxation, and, also, exciton transfer, and the radiative 
recombination processes has been applied to several 1.5-$nm$ semiconductor chalcogenide 
nanocrystals, such as Janus-type $Cd_{37}Pb_{31}Se_{68},~Cd_{31}Pb_{37}Se_{68},$ and to $Pb_{68}Se_{68}.$
Predictions for PL spectrum have been made, which were found to be in qualitative agreement 
with the existing experimental data.

In order to improve accuracy of this approach one should include computing corrections to the 
single-particle energies using $G_0W_0$ or DFT-based lattice field theory method
\cite{PhysRevResearch.3.023173}. Also, other relevant time-evolution channels, such 
as non-radiative recombination and carrier multiplication need to be included into BE. 
This is left to future work.

\section{Acknowledgment}
We are grateful to the group of Prof.~Kilina (NDSU) for providing us with atomistic models of
Janus NCs and to Prof. Kilin (NDSU) for suggesting PL spectrum calculation
based on BE.
The authors acknowledge use of computational resources of the Center 
for Computationally Assisted Science and Technology (CCAST) at North 
Dakota State University, 
%and NERSC No. DE-AC02-05CH11231, allocation Award 86185 for providing computational resources,
% Sveta's NSF proposal CHE-1413614
and financial support from the NSF grant CHE-2004197.

%\begin{thebibliography}{99}
%\end{thebibliography}
%\bibliographystyle{jpc}
%\bibliography{dftNSF2013}% Produces the bibliography via BibTeX.

%\bibliographystyle{unsrt}%if you want article's title; not specifying \bibliography at all is fine otherwise
%\bibliographystyle{plain}
\bibliography{dftNSF2020}%.bib Produces the bibliography via BibTeX.

% Override the revtex href command in order that the JHEP bib style
% will work properly:
\renewcommand{\href}[2]{#2} 

\end{document}